# The degradation of MgB$_2$ under ambient environment

A. Serquis[a], Y. T. Zhu, D. E. Peterson, and F. M. Mueller

Superconductivity Technology Center, MS K763, Los Alamos National Laboratory, Los Alamos, NM 87545, USA

R. K. Schulze

Materials Technology Metallurgy Group, MS G755, Los Alamos National Laboratory, Los Alamos, NM 87545, USA

V. F. Nesterenko and S. S. Indrakanti

Department of Mechanical and Aerospace Engineering, University of California, San Diego, La Jolla, CA 92093

The superconductivities of samples prepared by several procedures were found to degrade under ambient environment. The degradation mechanism was studied by measuring the change of surface chemical composition of dense MgB$_2$ pellets (prepared by hot isostatic pressure, HIPed) under atmospheric exposure using X-ray Photoelectron Spectroscopy (XPS). Results showed that samples with poor connectivity between grains and with smaller grain sizes degrade with time when exposed to ambient conditions. In these samples, the Tc did not change with time, but the superconducting transition became broader and the Meissner fraction decreased. In contrast, our well-sintered and the HIPed samples remained stable for several months under ambient condition. The degradation was found to be related to surface decomposition as observed by XPS. We observed the formation of oxidized Mg, primarily in the form of a Mg hydroxide, the increase of C and O contents, and the reduction of B concentration in the surface layer of MgB$_2$ samples.

---

[a] Electronic mail: aserquis@lanl.gov

The recent discovery of superconductivity in $MgB_2$ [1] has renewed the interest in studying its structural, magnetic and transport properties. Considerable progress has been made in understanding properties of this material and possibilities for its applications. However, to make practical devices, it is essential to understand the stability of $MgB_2$ under service environment. Most high-temperature superconductors are highly sensitive to moisture in air. Zhai et al.[2] studied the degradation of superconducting properties in $MgB_2$ films by exposing it to water and observed that the Tc(onset) of the films remained unchanged throughout the degradation process. There are so far no systematic study on the chemical stability of $MgB_2$, more specifically, its sensitivity to $O_2$, $H_2O$ and $CO_2$, under ambient conditions. The only observations about the $MgB_2$ reaction with air were the presence of surface boron oxides, identified by soft x-ray spectroscopy,[3] and magnesium oxide layers covering the grains of $MgB_2$, found using transmission electron microscopy.[4]

In this letter, we report the degradation under ambient conditions of $MgB_2$ samples prepared by different processes. As a result of exposure to air, the superconducting transition of samples with smaller grain sizes and poorly sintered became broader and the Meissner fraction decreased with time, while well-sintered samples or with larger grains remained largely unchanged. The degradation mechanisms were studied using X-ray Photoelectron Spectroscopy (XPS) on dense $MgB_2$ pellets prepared using Hot-isostatic pressing (HIP).

Four different samples were prepared by varying synthesis parameters (see Table I). Amorphous boron powder (-325 mesh, 99.99% Alfa Aesar) and magnesium turnings (99.98% Puratronic) were used as starting materials for all samples. For samples A and B, an atomic ratio of Mg: B = 1:1 was used. The boron powder was pressed into small pellets and all materials were wrapped in Ta foil. Sample A was placed in an alumina crucible inside a tube furnace under ultra-high purity Ar and heated at 900°C, as described elsewhere.[5] Sample B was sealed in a quartz tube under Ar atmosphere, heated at 900°C, slowly cooled down at 0.5°C/min to 600°C, and then fast cooled with water. For comparison, Sample C was prepared starting with a stoichiometric mixture, following a procedure of other authors,[6] in which Mg and B were mixed in an atomic ratio of Mg:B=1:2 without pressing pellets. The mixture was sealed in a Ta tube under Ar atmosphere, which was encapsulated in a quartz ampoule and heated for two hours at 950 °C, and then cooled to room temperature. Sample A was ground into a powder, from which Sample D was prepared by hot isostatic pressing (HIPing) at 200 MPa and 1000°C as previously reported. [7,8,9]

Powder X-ray diffraction data, collected using a Scintag XDS2000 θ−θ powder diffractometer, indicated that all samples were nearly single phase, with small amounts of MgO. The morphology of the samples was observed using a JEOL 6300FX scanning electron microscope (SEM). The susceptibility of the samples was measured using a superconducting quantum interference device (SQUID) magnetometer at 10 Oe. All as-synthesized samples have about the same Tc. However, their widths of superconducting transition and the grain sizes are quite different (see Table I).

To study their stability under ambient condition, we measured the same four samples (A, B, C and D) after exposing them to air for several periods of time. In Fig. 1 are plotted the dc magnetization vs. temperature for the $MgB_2$ samples A and B after preparation and after ageing. It can be seen that Sample A, which was prepared with excess Mg and had larger grain sizes, remained stable for several months under ambient conditions. The same behavior is observed in sample D. In contrast, sample B underwent a slow degradation. Although the Tc did not change, the superconducting fraction at 10 K was reduced by ~13% and the superconducting transition became broader ($\Delta Tc = 15.3$ K). The behavior of sample C was similar to sample B, with a reduction of the superconducting fraction by ~10 %. We did not observe any change in the XRD patterns of samples A, C and D, but Sample B showed a decrease in the intensity of $MgB_2$ phase and the presence of a very broad peak (in the 30-40 $2\theta$ region), which corresponds to the formation of an amorphous phase (see the inset in Fig. 2).

To study the nature and mechanisms of $MgB_2$ degradation, we performed X-ray Photoelectron Spectroscopy (XPS) only on sample D prepared by HIP, which can produce fully dense bulk samples suitable for surface analysis. All high-resolution XPS spectra were collected with a Physical Electronics 5600ci XPS system equipped with an Omni IV hemispherical capacitor. Standard samples of freshly cleaved MgO, natural brucite ($Mg(OH)_2$), and monolithic $MgCO_3$ were also examined under the same conditions. Atomic concentrations were calculated for the XPS spectra using the following relative sensitivity factors determined for the XPS system: B1s 0.171, C1s 0.314, O1s 0.733, and Mg2p 0.167.

The surface cleaned by Ar ion sputtering (estimated 300 Å of material removed) was taken as the intrinsic state, hereafter referred to as the initial one. XPS shows that the initial $MgB_2$ still has residual O and C present. The initial Mg-to-B ratio is lower than 1:2, which may indicate the presence of Mg vacancies in the external layer of the samples.

The relative atomic concentrations of Mg, B, C, and O as a function of exposure time are shown in figure 2. Most of compositional changes took place within the first 12 hours of exposure to air. The surface layer appeared to reach a constant composition in approximately 400 hours of air exposure. The relative Mg content in the surface layer dropped only slightly at a short exposure time (5 minutes), and then remained relatively constant. The B content, however, dropped rapidly and the O and C amounts climbed rapidly at short exposure times. This suggests that B in the $MgB_2$ was displaced by O and C containing species through reaction with $O_2$, $CO_2$ (and possibly also other C containing species), and $H_2O$ in air. The increase in C content was also observed as a growth in the intensity of the C1s peak with air exposure, and the addition of oxidized C (but not fully $MgCO_3$ like) after longer exposures. Likewise, the O1s peak grew in intensity with increasing air exposure, and there was a shift to higher binding energy from 531.1 eV to 532.0 eV. The binding energies for sample D and the standards are summarized in table II. We also observed a relatively broad O1s feature that suggests the presence of several chemically distinct oxygen species, in agreement with the observation of $B_2O_3$ and MgO by other authors.[3,4] The Mg2p photoemission line, which has a relatively insensitive binding energy with chemical environment,[10] showed only a slight shift to higher binding energy with air exposure. Likewise, the B1s photoemission line showed only slight increases in binding energy with increasing air exposure. In both the Mg2p and B1s cases, the greatest changes in chemical state took place within the first 12 hours of exposure.

In addition to the photoemission lines, the x-ray excited Mg KLL Auger transition was monitored. With increasing air exposure, the low energy feature at 302.1 eV, which is the signature of the intrinsic $MgB_2$, dropped in intensity, while the feature at 306.6 eV, which we believe to be the signature of oxidized Mg, increased in intensity (see Fig 3(a)). In order to sort out the specific chemical state of this oxidized Mg, several standards were examined. Figure 3(b) shows the Mg KLL spectra for MgO, $MgCO_3$, $Mg(OH)_2$, and spectrums acquired from the initial sample (sputter cleaned) and that exposed to air 1278 hours, hereafter referred to as air-exposed. Of the three oxidized Mg standards, the $Mg(OH)_2$ most closely matches the peak location in the air-exposed spectrum and the energy of the Mg2p photoemission line (see Table II). Therefore, we believe that the major degradation product in the $MgB_2$ from air exposure is $Mg(OH)_x$. However, it is also possible that the oxidized magnesium is in a mixed chemical environment, having some components of MgO and Mg oxycarbon in addition to the Mg hydroxide. Thus, a Mg oxy-hydroxide layer was formed on the surface of the sample. It is also possible that Mg oxy-hydroxide was formed at the grain boundaries in polycrystalline samples.

Assuming that Mg oxy-hydroxide overlayer on the surface of the air-exposed sample was homogeneous, the layer thickness was calculated as ~20 Å. This value was calculated from the ratio of the MgKLL line intensity contribution from the intrinsic $MgB_2$ lineshape and the contribution from the Mg hydroxide intensity from the $Mg(OH)_2$ standard lineshape.[11] Assuming a spherical grain, the percentage of oxy-hydroxide is only 0.1% for a grain with a diameter of 5 μm, but 6 % for a grain with a radius of 0.1 μm. Therefore, samples with large grain sizes or well-sintered, (for example Sample A and the HIPed sample) will not be affected by these surface reactions. However, a significant fraction of the samples will degrade in samples with smaller grain sizes (e.g. samples B and C). Schmidt et al.[12] suggested that chemical modifications of the surface (e.g. $MgB_2$ reaction with water) could potentially produce a layer with a lower Tc. These authors also suggested that if Mg vacancies are present in the near-surface regions, $H^+$ could enter the vacant $Mg^{2+}$ positions in the $MgB_2$ structure, or possibly form MgO.[11] To investigate if the Meissner degradation effect was related to Mg vacancies in $MgB_2$ samples, we determined the Mg occupancy by Rietveld analysis of powder X-ray data of these samples using the computer program GSAS.[13] It was found that all these samples had Mg vacancies lower than 0.01, much lower than 5% reported previously in other samples.[5] However, it is necessary to point out that XRD analysis can not accurately determine the Mg vacancies in the surface layer. The Mg excess in the initial formulation of the samples could help to prevent the presence of Mg vacancies, and might also have played a role in the formation of well-sintered samples of high quality.

In summary, the degradation of $MgB_2$ in air was caused by a surface decomposition as observed by XPS. Upon air exposure, the surfaces and possibly also the interfaces oxidized, and more specifically in this case, hydroxylated, to form magnesium oxy-hydroxide. Samples with a small grain sizes and not well-sintered are more susceptible to degradation when exposed to air. High-quality stable samples can be synthesized using our method, which requires the presence of Mg excess in an open atmosphere under Ar flux.